\documentclass[aps,prl,twocolumn,floatfix,showpacs,citeautoscript,longbibliography,hyperlinks,superscriptaddress]{revtex4-2}
\usepackage{amsmath}
\usepackage{amssymb}
\usepackage{amsfonts}
\usepackage{dsfont}
\usepackage{color}
\usepackage{nicefrac}
\usepackage[autostyle]{csquotes}
\usepackage[colorlinks=true,citecolor=blue]{hyperref}
\usepackage{graphicx,graphics}

\newcommand{\revision}[1]{{{#1}}}
\newcommand{\revisiontwo}[1]{{{#1}}}
\newcommand{\revisionthree}[1]{{{#1}}}

\begin{document}

\title{Lee-Yang theory of criticality in interacting quantum many-body systems}

\author{Timo Kist}
\affiliation{Department of Applied Physics, Aalto University, 00076 Aalto, Finland}
\affiliation{Fakult{\"a}t f{\"u}r Physik, Georg-August-Universit{\"a}t G{\"o}ttingen, 37077 G{\"o}ttingen, Germany}
\author{Jose L. Lado}
\affiliation{Department of Applied Physics, Aalto University, 00076 Aalto, Finland}
\author{Christian Flindt}
\affiliation{Department of Applied Physics, Aalto University, 00076 Aalto, Finland}

\begin{abstract}
Quantum phase transitions are a ubiquitous many-body phenomenon that occurs in a wide range of physical systems, including superconductors, quantum spin liquids, and topological materials. However, investigations of quantum critical systems also represent one of the most challenging problems in physics, since highly correlated many-body systems rarely allow for an analytic and tractable description. Here we present a Lee-Yang theory of quantum phase transitions including a method to determine quantum critical points which readily can be implemented within the tensor network formalism and even in realistic experimental setups. We apply our method to a quantum Ising chain and the anisotropic quantum Heisenberg model and show how the critical behavior can be predicted by calculating or measuring the high cumulants of properly defined operators. Our approach provides a powerful
formalism to analyze quantum phase transitions using tensor networks, and it paves the way for systematic investigations of quantum criticality in two-dimensional systems.
\end{abstract}

\date{\today}

\maketitle

\emph{Introduction.}--- Predicting and understanding the phase behavior of correlated quantum many-body systems constitute one of the most demanding problems in theoretical condensed matter physics and related fields \cite{Sachdev2000,Vojta2003}. Of particular importance is the identification of critical points in parameter space, which separate quantum many-body ground states that have fundamentally different properties on macroscopic length scales. Current efforts relate to high-$T_c$ superconductivity \cite{PhysRevX.10.031016}, quantum magnetism \cite{Schollwock2004}, and designer materials \cite{PhysRevX.8.011044}, among many others. Progress, however, is hindered by the fact that correlated many-body systems are rarely analytically tractable, and computational approaches are typically demanding, in particular for systems in dimensions higher than one. The entanglement in a quantum many-body system is one sensitive probe of criticality \cite{Osterloh2002,Lambert2004}, still, finding a unifying and convenient framework to describe quantum phase transitions remains a challenge.

A powerful method that has proven successful in describing thermodynamic phase transitions is the Lee-Yang formalism, which considers the zeros of the partition function in the complex plane of the external control parameters \cite{Yang1952,Lee1952,Blythe2003,Bena2005}. With increasing system size, the complex partition function zeros move towards the points on the real-axis, where a phase transition occurs. The Lee-Yang formalism has in recent years experienced a surge of interest because of several experiments that have determined the partition function zeros in a variety of physical systems \cite{PhysRevLett.81.5644,PhysRevLett.114.010601,Brandner2017,francis2020} and thereby shown that Lee-Yang zeros are not just a theoretical concept. In fact, they provide an efficient tool to predict and understand phase transitions in interacting many-body systems, also experimentally \cite{Flindt2013,Deger2018,Deger2019,Deger2020a,Deger2020b}. However, while the focus so far has been on classical systems, a Lee-Yang theory to describe quantum phase transitions is now clearly becoming desirable~\footnote~.

\begin{figure}[b!]
  \centering
  \includegraphics[width=0.98\columnwidth]{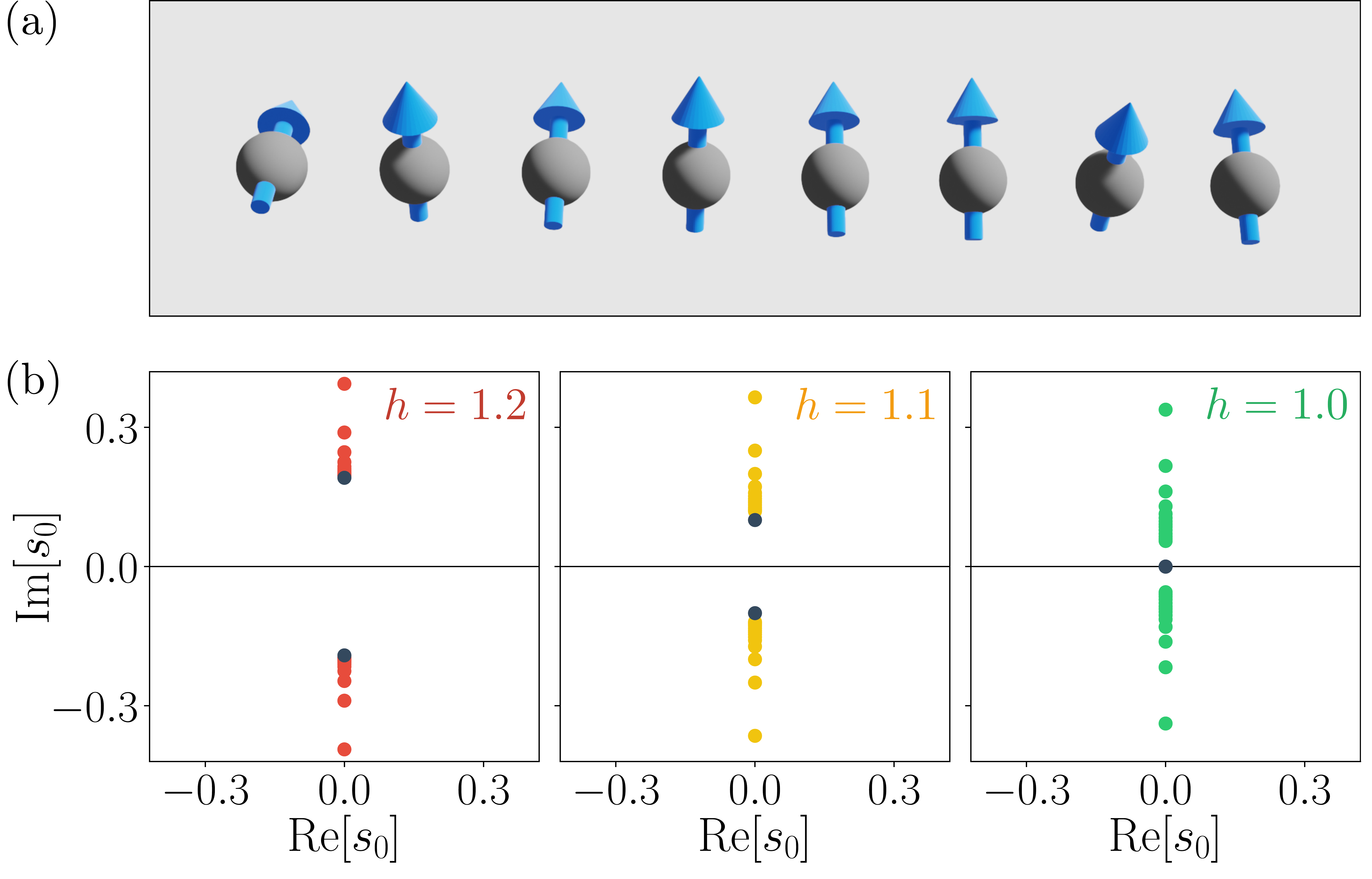}
  \caption{Quantum Ising chain and Lee-Yang zeros. (a) The quantum Ising chain consists of $L$ interacting quantum spin-$\frac{1}{2}$ particles in a transverse magnetic field, $h$. (b) Complex Lee-Yang zeros of the moment generating function with increasing chain length and their convergence points (in black) in the thermodynamic limit obtained by extrapolation in Fig.~\ref{fig:fig2}(a). For $h=1.2$ and $h=1.1$, the Lee-Yang zeros remain complex in the thermodynamic limit, while for $h=1.0$, they reach the real-axis at $s=0$, signaling a quantum phase transition.}
  \label{fig:LY_zeros}
\end{figure}

In this paper, we present a Lee-Yang theory of quantum phase transitions, which extends the Lee-Yang description of thermodynamic phase transitions to the quantum realm. At zero temperature, the partition function reduces to just a single Boltzmann factor, and it is not clear how to apply the standard Lee-Yang formalism. Here, we suggest instead that one consider the fluctuations of a properly chosen observable together with the complex zeros of the associated moment generating function. The idea is illustrated in Fig.~\ref{fig:LY_zeros}, showing a quantum Ising chain and the complex zeros of the moment generating function of the magnetization, which we determine from the high magnetization cumulants as we will see. At magnetic fields above the critical value, the Lee-Yang zeros remain complex in the thermodynamic limit. By contrast, at criticality, they reach the real-axis, signaling a quantum phase transition. Below, we apply our method both to the quantum Ising chain and the anisotropic quantum Heisenberg model for which we find the high cumulants using DMRG calculations \cite{White1992,PhysRevB.48.10345,Schollwck2011,ITensor,dmrgpy,2020arXiv200714822F}. We also discuss a connection between the Lee-Yang zeros and the large-deviation statistics of rare fluctuations \cite{Touchette2009}, \revision{which characterize the states on each side of the critical point.} Our Lee-Yang method can detect emergent criticality in relatively short chains, making the approach attractive for situations where the system size may be a limiting factor. Among others, those include numerical calculations \cite{Yan2011,PhysRevX.8.011044,Hu2019,Jiang2019,PhysRevX.10.031016}, measurements on nano-structures~\cite{Choi2019} or atoms in optical lattices~\cite{Eckardt2017}, or even near-term quantum computations~\cite{CerveraLierta2018}. Moreover, since the Lee-Yang zeros can be expressed in terms of simple
expectation values, the formalism is suitable for a variety of tensor \cite{White1992,PhysRevB.48.10345,Verstraete2008,PhysRevLett.99.220405,Schollwck2011}
and neural network techniques \cite{Carleo2017,PhysRevLett.124.020503,PhysRevResearch.2.023358,Melko2019,PhysRevE.101.053301,PhysRevB.100.125124}.  

\emph{Quantum Ising chain.}--- To be specific, we first consider the transverse-field quantum Ising model in one dimension. \revision{However, it should become clear that our Lee-Yang method can be applied to many quantum phase transitions for which an order parameter can be defined, also in higher dimensions than one.} The quantum Ising model describes a chain of interacting quantum spin-$\frac{1}{2}$ particles governed by the Hamiltonian
\begin{align}
    \hat{H} = - \sum_{\ell=1}^L (h\hat{\sigma}_\ell^x + \hat{\sigma}_\ell^z \hat{\sigma}_{\ell+1}^z),
    \label{eq:H}
\end{align}
where $\hat{\sigma}_\ell^x$ and $\hat{\sigma}_\ell^z$ are the usual Pauli operators for the spin on site $\ell$, the transverse magnetic field (in suitable units) is denoted by $h$, and we consider periodic boundary conditions, $\hat{\sigma}_{L+1}^z=\hat{\sigma}_{1}^z$, where $L$ is the chain length that goes to infinity in the thermodynamic limit. Without a magnetic field, $h=0$, the ground state is a classical ferromagnetic state with a finite magnetization in the $z$-direction. On the other hand, at large magnetic
fields, all spins point in the $x$-direction, and the average magnetization vanishes. For long chains, a quantum order-disorder transition develops at $h=1$ with a continuous increase in the magnetization as the magnetic field is decreased~\cite{Pfeuty1970}.

To understand the phase behavior of the system, we consider the distribution of the order parameter \cite{Eisler2003,Cherng2007,Lamacraft2008,Ivanov2013,Groha2018,Collura2019,Xu2019},
\begin{equation}
    P_L(M_z) = \langle \Psi_0|\delta(M_z-\hat{M}_z)|\Psi_0 \rangle,
    \label{eq:kpm}
\end{equation}
where $\hat{M}_z = \sum_{\ell=1}^L \hat{\sigma}_\ell^z$ is the operator for the total magnetization, and $|\Psi_0 \rangle$ is the ground state of the Hamiltonian. We also define the moment generating function,
\begin{align}
    \chi_L(s) = \int_{-L}^L dM_z \; P_L(M_z) \; e^{sM_z}=\langle\Psi_0|e^{s\hat{M}_z}|\Psi_0\rangle, \label{eq:mgf}
\end{align}
and the cumulant generating function, $\Theta_L(s)=\ln[\chi_L(s)]$, which has a well-defined thermodynamic limit denoted as $\theta(s)\equiv\lim_{L\rightarrow\infty} \Theta_L(s)/L$. For classical systems, the cumulants of thermodynamic observables can be obtained by differentiating the free energy with respect to a conjugate field \cite{Deger2018,Deger2019,Deger2020a,Deger2020b}. By contrast, in our case, because of quantum fluctuations and non-commuting observables~\cite{Ivanov2013}, there is no simple relation between the cumulant generating function and the free energy, i.e.~the ground state energy at zero temperature, and the cumulants are defined as $ \langle\!\langle M_z^n \rangle\!\rangle = \partial_s^n\Theta_L(s)|_{s=0}.$ 

\emph{Lee-Yang zeros.}--- As in the classical Lee-Yang theory of phase transitions \cite{Yang1952,Lee1952,Blythe2003,Bena2005}, we now consider the complex zeros of the function that generates the moments of the order parameter. Hence, in the quantum case, we consider the moment generating function instead of the partition function as for classical systems. For finite chains, the moment generating function is a finite sum of exponentials, and it is thus an entire function, which can be factorized in terms of its complex zeros as \cite{Arfken2012}
\begin{align}
    \chi_L(s) = e^{c s}\prod_k\left(1-s/s_k\right),
\end{align}
where $c$ is a constant, and the zeros come in complex conjugate pairs, since the moment generating function is real for real $s$. As the system size is increased, the Lee-Yang zeros will move towards the points in the complex plane, where the cumulant generating function becomes non-analytic in the thermodynamic limit \cite{Yang1952,Lee1952,Blythe2003,Bena2005,Flindt2013,Deger2018,Deger2019,Deger2020a,Deger2020b}. The magnetization cumulants inherit this non-analytic behavior, and if the Lee-Yang zeros reach $s=0$, some of the cumulants will diverge in the thermodynamic limit, signaling a quantum phase transition.

\begin{figure*}
  \includegraphics[width=2\columnwidth]{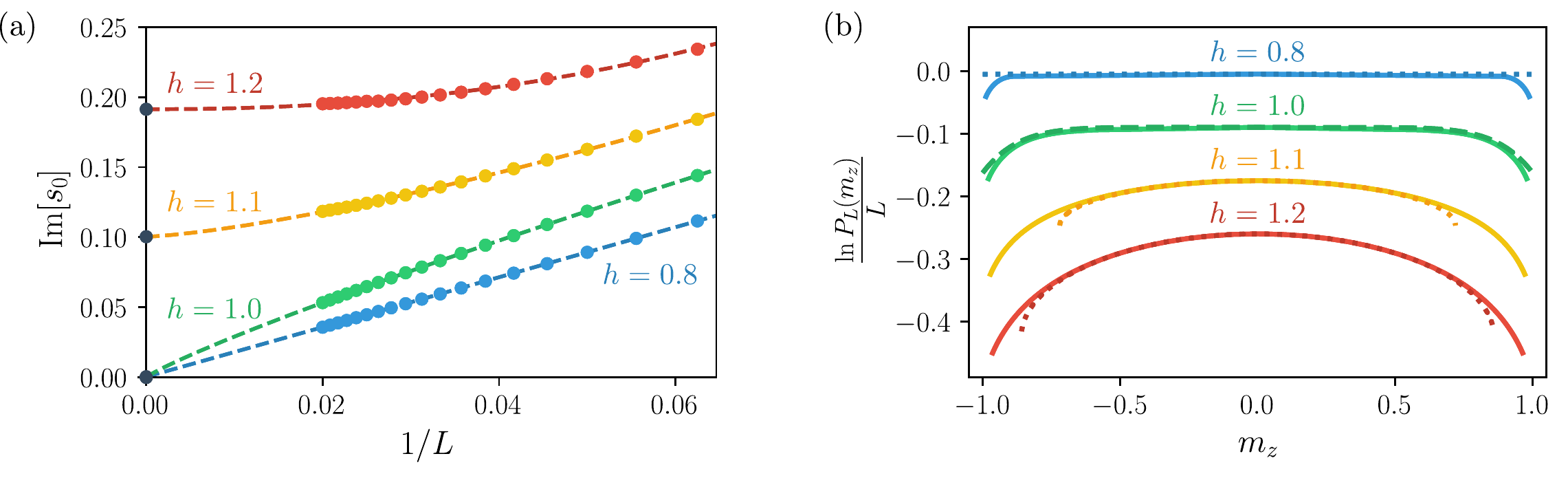}
  \caption{Scaling analysis and large-deviation statistics. (a) The leading Lee-Yang zeros are obtained from Eq.~(\ref{eq:LY_abs}) based on DMRG calculations of the magnetization cumulants of order $n=7,8,9,10$, and we have checked that the results are unchanged if we increase the cumulant order.  The real part of the Lee-Yang zeros is close to zero, while we extrapolate the convergence point of the imaginary part in the thermodynamic limit using the scaling ansatz in Eq.~(\ref{eq:scaling}). The convergence points, $s_c$, are shown with a black circle here and in Fig.~\ref{fig:LY_zeros}(b).  (b) Large-deviation statistics for the quantum Ising model calculated using DMRG together with the approximation in Eq.~(\ref{eq:LDF_LY}) with the convergence points of the Lee-Yang zeros for $h=0.8$, $h=1.1$, and $h=1.2$ inserted (dotted lines). For $h=1.0$, we show the analytic expression (dashed line) in Eq.~(\ref{eq:LDF_crit}), which fits well with our DMRG results. For the sake of clarity, we have vertically shifted the curves, which otherwise would reach zero at their maximum.
  }
  \label{fig:fig2}
\end{figure*}

To determine the motion of the Lee-Yang zeros in the complex plane, we turn to the cumulants of the magnetization in finite chains. Using the definition of the cumulants in combination with the factorization of the moment generating function, we can write them as 
\begin{align}
    \langle\!\langle M_z^n \rangle\!\rangle = - (n-1)! \sum_k 1/s_k^{n}, \,\, n>1,
    \label{eq:LYcumu}
\end{align}
showing how the cumulants are determined by the Lee-Yang zeros. Moreover, for high orders, the sum is dominated by the pair of zeros, $s_0$ and $s_0^*$, that are closest to $s=0$, while the relative contributions from other zeros are suppressed as $|s_0/s_k|^n$, and they can thus be neglected for sufficiently high orders. It is then possible to solve Eq.~(\ref{eq:LYcumu}) for the closest pair of zeros and express them in terms of four consecutive cumulants as~\cite{Flindt2013,Deger2018,Deger2019,Deger2020a,Deger2020b}
\begin{equation}
\begin{split}
    \textrm{Re}[s_0]\! &\simeq \! \frac{(n\!+\!1)\langle\!\langle M_z^{n}\rangle\!\rangle\!\langle\!\langle M_z^{n+1}\rangle\!\rangle\!-\!(n\!-\!1)\langle\!\langle M_z^{n-1}\rangle\!\rangle\!\langle\!\langle M_z^{n+2}\rangle\!\rangle}{2(1\!+\!1/n)\langle\!\langle M_z^{n+1}\rangle\!\rangle^{2}\!-\!2\langle\!\langle M_z^{n}\rangle\!\rangle\!\langle\!\langle M_z^{n+2}\rangle\!\rangle}, \\
    |s_0|^2 \! &\simeq \! \frac{n^{2}(n\!+\!1)\langle\!\langle M_z^{n}\rangle\!\rangle^{2}\!-\!n(n^{2}\!-\!1)\langle\!\langle M_z^{n-1}\rangle\!\rangle\!\langle\!\langle M_z^{n+1}\rangle\!\rangle}{(n\!+\!1)\langle\!\langle M_z^{n+1}\rangle\!\rangle^{2}\!-\!n\langle\!\langle M_z^{n}\rangle\!\rangle\!\langle\!\langle M_z^{n+2}\rangle\!\rangle}.
\end{split}
\label{eq:LY_abs}
\end{equation}
Importantly, by measuring or calculating the high cumulants of the magnetization, it is thus possible to determine the closest pair of Lee-Yang zeros and follow their motion as the system size is increased.  The fact that the Lee-Yang zeros can be written in terms of simple expectation values of operators makes the formalism suitable for a variety of tensor network methods developed for interacting quantum many-body systems~\cite{White1992,PhysRevB.48.10345,Verstraete2008,PhysRevLett.99.220405,Schollwck2011}. \revisiontwo{Moreover, any order parameter can be inserted in Eq.~(\ref{eq:LY_abs}) to determine the corresponding Lee-Yang zeros.}

\emph{Quantum phase transition.}--- We are now ready to illustrate our method on the quantum Ising model. To this end, we have made DMRG calculations of the high magnetization cumulants and subsequently determined the closest Lee-Yang zeros using Eq.~(\ref{eq:LY_abs}). Figure~\ref{fig:LY_zeros}(b) shows the extracted Lee-Yang zeros for three different magnetic fields and increasing chain length. For $h=1.2$ and $h=1.1$, the Lee-Yang zeros remain complex in the thermodynamic limit, however, they come closer to the real-axis as we lower the magnetic field, and for $h=1.0$, they reach $s=0$, signaling a quantum phase transition. To find the convergence points in the thermodynamic limit, we use the scaling ansatz~\cite{Deger2019,Deger2020a,Deger2020b}
\begin{equation}
\textrm{Im}[s_0-s_c]=  \alpha L^{-\gamma},
\label{eq:scaling}
\end{equation}
where the parameters $\alpha$ and $\gamma$ together with the convergence point $s_c$ are determined in Fig.~\ref{fig:fig2}(a) using least-squares fitting\revision{, leading to the estimate $h_c\simeq 1.0007(3)$ for the critical field. As for the classical Ising model in $d$ dimensions, we expect that the parameter $\gamma$ at criticality is given by $\gamma=d-g$, where $g$ is the critical exponent~\cite{Deger2020b}. From Fig.~\ref{fig:fig2}(a), we then find $g\simeq 0.1283(2)$, which is close to the exact value of $g=1/8$.  Moreover, we can predict the critical behavior of the system for short chains of length $L\sim20-40$.} In Fig.~\ref{fig:fig2}(a), we also see how the Lee-Yang zeros converge to $s=0$ below the critical magnetic field. In this case, the ground state has become twofold-degenerate, such that a small symmetry-breaking field would cause a first-order phase transition.  

\emph{Large-deviation statistics.}--- Having extracted the convergence points of the Lee-Yang zeros, we now discuss how they encode important information about the rare fluctuations of the magnetization\revisiontwo{, also away from the critical point}. According to the G{\"a}rtner-Ellis theorem~\cite{Touchette2009}, the large-deviation statistics of the magnetization can for long chains be obtained as $\ln [P_L(m_z)]/L \simeq \theta(s_m)-m_z s_m$,
where $m_z=M_z/L$ is the magnetization per site, and $s_m$ solves the saddle-point equation, $\theta'(s)-m_z=0$. The saddle-point equation can be rewritten in the appealing form $\langle \hat{m}_z\rangle (s) \equiv \langle \Psi_0(s)| \hat{m}_z |\Psi_0(s)\rangle=m_z$, where $\hat{m}_z=\hat{M}_z/L$ is the operator for the magnetization per site, and we have defined the (normalized) $s$-biased ground state, $|\Psi_0(s)\rangle= e^{s\hat{M}_z/2} |\Psi_0\rangle/\sqrt{\langle \Psi_0|e^{s\hat{M}_z} |\Psi_0\rangle}$. This rewriting resembles the $s$-ensemble formalism that has been used to investigate rare dynamical trajectories in glass formers \cite{Hedges2009} and in open quantum systems \cite{Garrahan2010}: A positive biasing field, $s>0$, enhances the parts of the ground state with positive magnetization, while for $s<0$ it enhances those with negative magnetization. Evaluating $P_L(m_z)$ by solving the saddle-point equation then amounts to finding the biasing field for which the average magnetization equals $m_z$, i.e.~$\langle \hat{m}_z\rangle(s)=m_z$. As such, large-deviation theory provides a window into the properties of the quantum many-body ground state beyond unbiased averages.

In Fig.~\ref{fig:fig2}(b), we show the large-deviation statistics for the quantum Ising model above, below, and at the critical point, obtained by numerically solving the saddle-point equation for sufficiently long chains. At the critical point and below, the large-deviation function is mostly flat, except for the tails close to full magnetization. By contrast, above the critical point, the large-deviation function takes on a curved form, whose shape can be understood by considering the Lee-Yang zeros. For classical Ising models and several other equilibrium systems, it has recently been argued that the large-deviation function can be rather well approximated using only the convergence points of the Lee-Yang zeros. If the real part of the convergence points vanishes, the large-deviation function can be approximated by the ellipse \cite{Brandner2017,Deger2018,Deger2019,Deger2020a,Deger2020b}
\begin{align}
    \frac{\ln P_L(m_z)}{L} \simeq |\textrm{Im}[s_c]|\left(\sqrt{\bar{m}_z^2-m_z^2}-\bar{m}_z\right) \label{eq:LDF_LY},
\end{align}
where $\bar{m}_z$ is a  free parameter close to one. Below and at the critical point, the Lee-Yang zeros converge to zero, and we expect a flat large-deviation function, as we also find in Fig.~\ref{fig:fig2}(b). Above the critical point, where the Lee-Yang zeros remain complex, the large-deviation statistics are also well-captured by Eq.~(\ref{eq:LDF_LY}) as we see in Fig.~\ref{fig:fig2}(b) and as we have checked for higher magnetic fields. \revision{Importantly, the large-deviation statistics allow us to clearly distinguish the states on each side of the critical point.}

While Eq.~(\ref{eq:LDF_LY}) provides a general but somewhat crude approximation of the large-deviation statistics, we can obtain a more accurate expression at criticality using an approximation for the moment generating function found by Lamacraft and Fendley~\cite{Lamacraft2008}. Using their result, the scaled cumulant generating function becomes $\theta(s)\simeq |m_0 s|^{8/7}$, where the constant $m_0=\Gamma(7/8)[2\sqrt{\pi}\Gamma(3/7)/\Gamma(13/14)]^{7/8}/(2\pi)\simeq0.952$ is given in terms of the gamma function $\Gamma(x)$. By solving the saddle-point equation, we then arrive at a remarkably simple large-deviation function for the Ising chain at criticality,
\begin{equation}
    \frac{\ln P_L(m_z)}{L} \simeq -\frac{7^7}{8^8}\left(\frac{m_z}{m_0}\right)^8,\,\, h=1, \label{eq:LDF_crit}
\end{equation}
which indeed agrees well with our numerics in Fig.~\ref{fig:fig2}(b).

\emph{Anisotropic Heisenberg chain.}--- As the second application of our method, we turn to the one-dimensional Heisenberg model, where each site of the chain is now a spin-$1$ particle, and the Hamiltonian reads~\cite{PhysRevB.67.104401,PhysRevLett.50.1153,PhysRevB.79.054412}
\begin{equation}
    \hat{H} = \sum^L_{\ell=1} \left ( \hat{S}_\ell^x \hat{S}_{\ell+1}^x + \hat{S}_\ell^y \hat{S}_{\ell+1}^y + J_z \hat{S}_\ell^z \hat{S}_{\ell+1}^z + D (\hat{S}_\ell^z)^2 \right ).
\end{equation}
In the isotropic limit ($J_z=1$, $D=0$), the system has a quantum-disordered ground state with topological order, giving rise
to fractional edge excitations \cite{PhysRevLett.50.1153}. In addition, for large values of $D$ and $J_z$, it maps to a classical
Ising antiferromagnet, whose ground state is a time-reversal symmetry broken antiferromagnet \cite{PhysRevB.79.054412}. We now construct the phase boundary between the N\'{e}el phase and the Haldane phase of the system using our Lee-Yang method. The order parameter is given by the staggered magnetization $\hat{M}^{S}_z = \sum_{\ell=1}^L (-1)^\ell \hat{S}_\ell^z$, whose average vanishes in the Haldane phase, and is finite in the N\'{e}el  phase, and it enters in the cumulant averages in Eq.~(\ref{eq:LY_abs}).

\begin{figure}
  \includegraphics[width=.98\columnwidth]{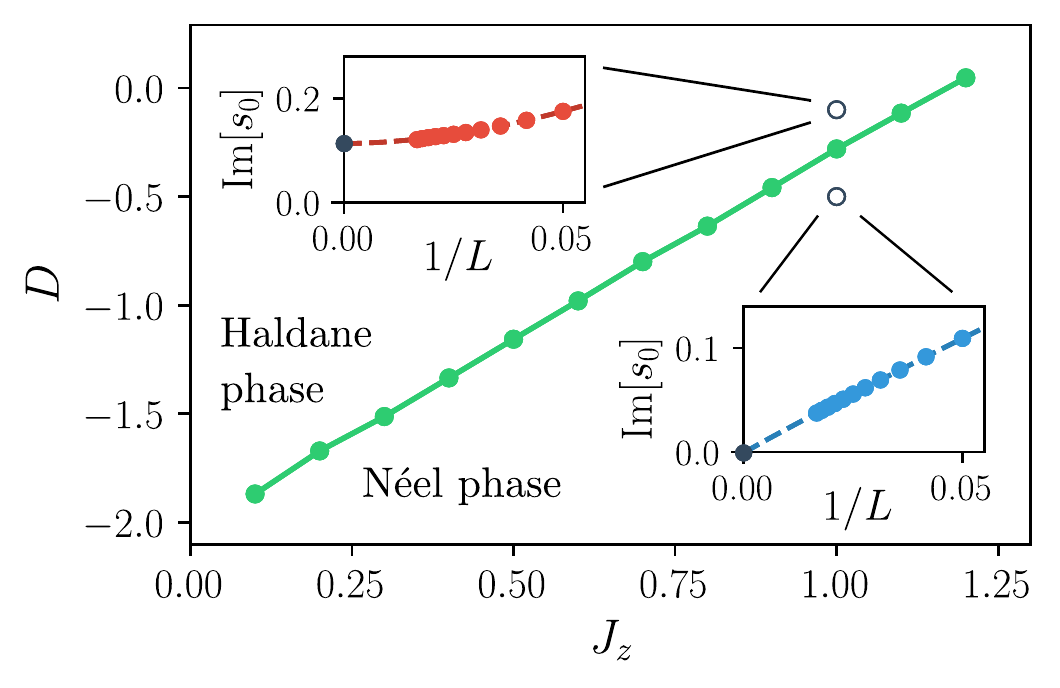}
  \caption{The anisotropic Heisenberg chain. The green line indicates the phase boundary between the Haldane and the N\'{e}el phase obtained with our cumulant method. In the Haldane phase, the Lee-Yang zeros remain complex, while they converge to $s=0$ in the N\'{e}el phase, similar to the quantum Ising chain above and below the critical point. The insets illustrate the extrapolation of the convergence points in the two phases, for  $D = -0.1$ and $D = -0.5$, respectively, and $J_z = 1.0$.}
  \label{fig:fig3}
\end{figure}

Figure~\ref{fig:fig3} shows the phase boundary that we have identified from the convergence points of the Lee-Yang zeros. In the Haldane phase, the Lee-Yang zeros remain complex in the thermodynamic limit, while they reach the real-axis in the N\'{e}el phase, as illustrated in the insets, where we show the determination of the convergence points in the thermodynamic limit for two points in the phase diagram. Just as for the quantum Ising chain below the critical point, the ground state of the N\'{e}el phase is twofold degenerate, such that a small symmetry-breaking field would cause a first-order phase transition.

Finally, we have also applied our method to \revisionthree{the quantum Ising lattice~\cite{Blote2002} and the $J_1$--$J_2$ model~\cite{PhysRevB.100.125124,Gong2014} in two dimensions and found phase transitions in these systems using small lattices of up to size $L\times L = 4\times4$~\cite{Note2}.  }

\emph{Conclusions.}--- We have presented a Lee-Yang theory of quantum phase transitions that is based on the complex zeros of moment generating functions. The critical behavior of a quantum many-body system can then be predicted from the high cumulants of suitable  observables. Importantly, the method only requires calculations of ground state expectation values, making it attractive for a variety of tensor and neural network techniques. We have illustrated the method using DMRG calculations for a quantum Ising chain and the anisotropic Heisenberg model and have discussed the connection between the Lee-Yang zeros and the large-deviation statistics of rare fluctuations. We have also \revisionthree{applied our method to two two-dimensional quantum systems, indicating that it can be applied to a larger class of systems beyond one dimension.} In addition, by using thermal averages for the moments and cumulants, it may be possible to implement the method at finite temperatures. \revision{Finally, for systems whose degrees of freedom map to free fermions with conserved total particle number, the entanglement entropy can be expressed as a series in the cumulants of the bipartite fluctuations \cite{song11,song12}, suggesting that a connection between Lee-Yang zeros and the entanglement may also exist, at least for some systems.}

\emph{Acknowledgements.}--- We thank A.~Deger and F.~Brange for useful discussions and  acknowledge the computational resources provided by the Aalto Science-IT project. We acknowledge support from the Academy of Finland through the Finnish Centre of Excellence in Quantum Technology (project numbers 312057 and 312299) and grants number 308515, 331342, and 336243.

%\bibliography{biblio}

%

\end{document}

\title{Supplemental material: \\ Lee-Yang Theory of Criticality in Interacting Quantum Many-Body Systems}

\author{Timo Kist}
\affiliation{Department of Applied Physics, Aalto University, 00076 Aalto, Finland}
\affiliation{Fakult{\"a}t f{\"u}r Physik, Georg-August-Universit{\"a}t G{\"o}ttingen, 37077 G{\"o}ttingen, Germany}
\author{Jose L. Lado}
\affiliation{Department of Applied Physics, Aalto University, 00076 Aalto, Finland}
\author{Christian Flindt}
\affiliation{Department of Applied Physics, Aalto University, 00076 Aalto, Finland}

\maketitle

\begin{figure*}
	\includegraphics[width=0.95\columnwidth]{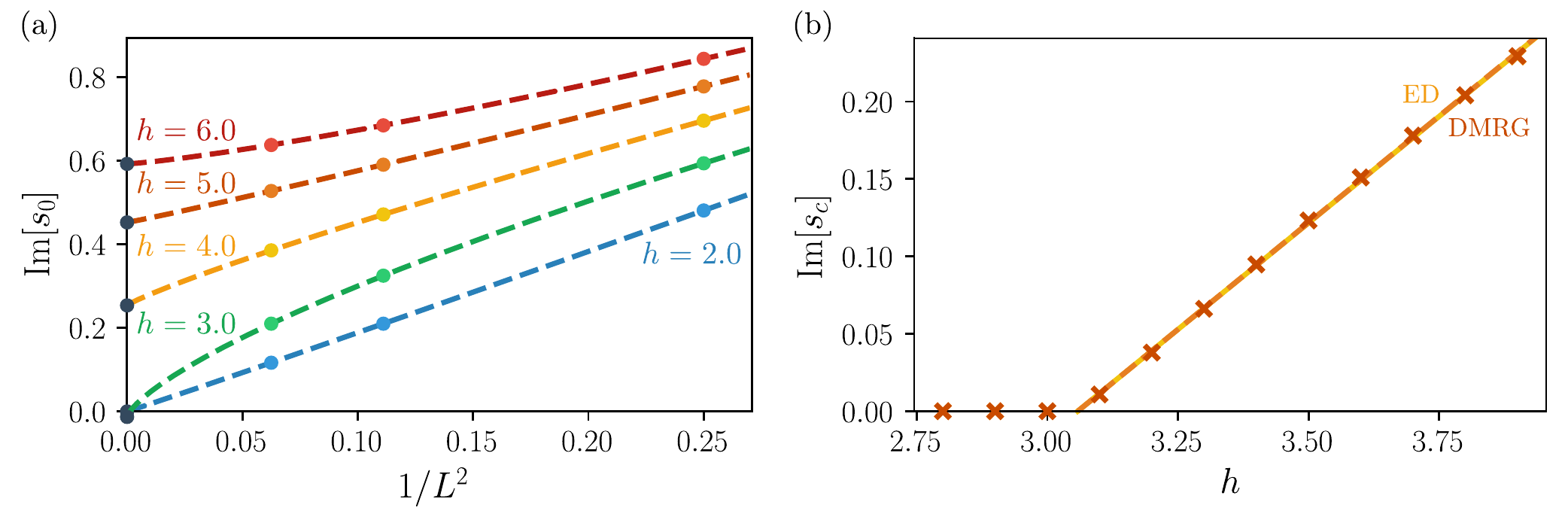}
	\caption{Two-dimensional quantum Ising model. (a) Finite-size scaling of the extracted Lee-Yang zeros for different magnetic fields. (b) Lee-Yang zeros in the thermodynamic limit as a function of the magnetic field. We find $h_c\simeq 3.06$ with $J=1$ \cite{Blote2002}.}
	\label{fig:SM1}
\end{figure*}

\emph{Two-dimensional quantum Ising model.}--- The Hamiltonian of the  quantum  Ising model on a square lattice reads
\begin{equation}
	\hat{H} =
	-J \sum_{\langle i,j \rangle} \hat{\sigma}^z_i \hat{\sigma}^z_j 
	-
	h \sum_i \hat{\sigma}_i^x,
\end{equation}
where $\langle i,j \rangle$ denotes summation over nearest neighbors, the indices run over all sites of the lattice of size $L\times L$ with periodic boundary conditions,
and $\hat{\sigma}^x_i$ and $\hat{\sigma}^z_i$ are the usual Pauli operators. In Fig.~\ref{fig:SM1}, we show the determination of the critical point using lattices of sizes up to $L\times L=4\times 4$ using DMRG and exact diagonalization (ED). Panel (a) shows the finite-size scaling analysis of the Lee-Yang zeros for different magnetic fields, and panel (b) shows the convergence points in the thermodynamic limit as a function of the magnetic field. We find a critical field of $h_c\simeq 3.06\,J$~\cite{Blote2002}.

\begin{figure*}
	\includegraphics[width=0.95\columnwidth]{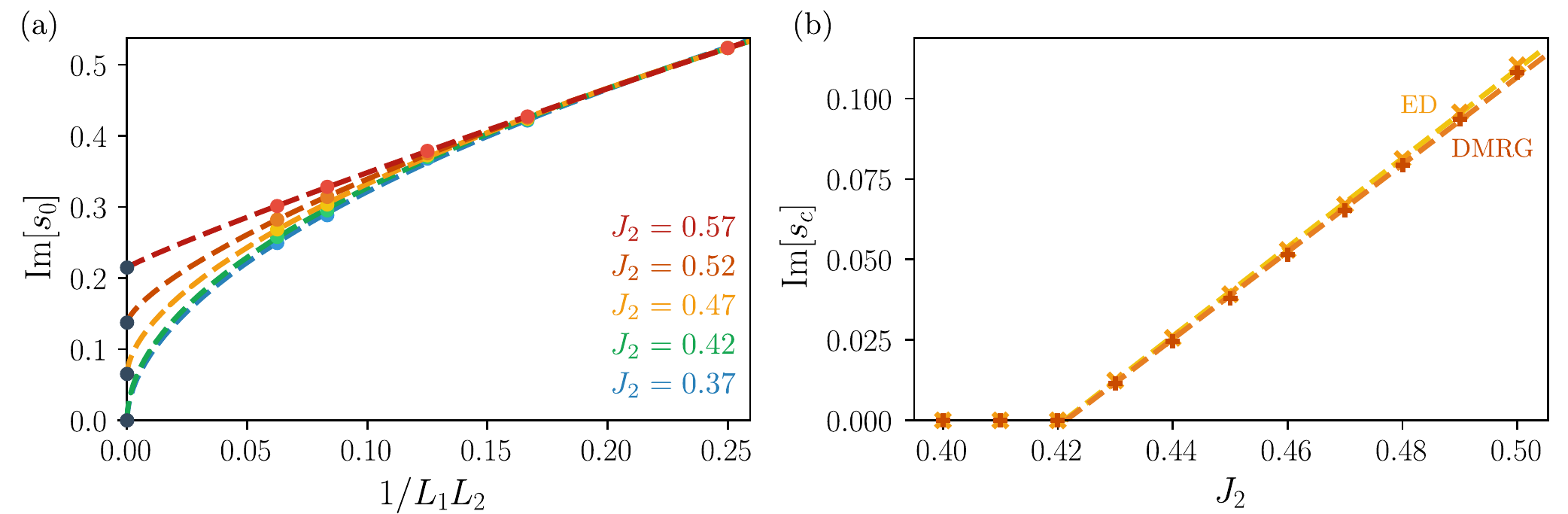}
	\caption{Two-dimensional $J_1$--$J_2$ model. (a) Finite-size scaling of the extracted Lee-Yang zeros for different couplings. (b) Lee-Yang zeros in the thermodynamic limit as a function of the coupling. We find a critical coupling of $J_2 \simeq 0.42$ with $J_1=1$~\cite{Gong2014,PhysRevB.100.125124}.
	}
	\label{fig:SM2}
\end{figure*}

\vspace{0.25cm}
\emph{Two-dimensional $J_1$--$J_2$ model.}--- The Hamiltonian of the  $J_1$--$J_2$ model on a rectangular lattice reads
\begin{equation}
	\hat{H} =
	J_1 \sum_{\langle i,j \rangle} \vec S_i \cdot \vec S_j 
	+
	J_2 \sum_{\langle\! \langle i,j \rangle\! \rangle} \vec S_i \cdot \vec S_j 
\end{equation}
where $\langle i,j \rangle$ and $\langle\! \langle i,j \rangle\! \rangle$ denote summation over nearest and next-nearest neighbors, respectively, the indices run over all sites of the lattice of size $L_1\times L_2$ with open boundary conditions, and
$\vec S_i = (\hat{S}^x_i,\hat{S}^y_i,\hat{S}^z_i)$ are the local
spin-$\frac{1}{2} $ operators.  In Fig.~\ref{fig:SM2}, we show the determination of a quantum phase transition between the N\'{e}el phase and the spin-liquid phase using lattices of sizes up to $L_1\times L_2=4\times 4$. Panel (a) shows the finite-size scaling analysis of the Lee-Yang zeros for different values of the coupling, $J_2$, and panel (b) shows the convergence points in the thermodynamic limit as a function of the coupling. We find a critical coupling of $J_2 \simeq 0.42\,J_1$~\cite{Gong2014,PhysRevB.100.125124}

